\documentclass{article}
\usepackage{epsfig}
\usepackage{amstex}
\usepackage{amssymb}
\usepackage{sprocl}
\newcommand{\psibar} {\bar{\psi}}

\newcommand{\mq} {\ensuremath{m_{\text{q}}}}

\newlength{\hepwidth}
\begin{document}

\title{\hfill\parbox{\hepwidth}{
\textnormal{hep-lat/9701009}}\\
Faster Fermions in the Tempered Hybrid Monte Carlo 
	Algorithm\footnote{Talk presented at the conference
	{\em Multi-Scale Phenomena and their Simulation} from
	30.9.-4.10 1996 in Bielefeld
      }}

\author{ G. BOYD }

\address{ Dipartimento di Fisica dell'Universit\`a,
    I-56126 Pisa, Italy \\
After 1. Nov. 1996: Center for Computational Physics, 
University of Tsukuba, 
Tsukuba, Ibaraki 305, Japan. ~~~
email:boyd@@rccp.tsukuba.ac.jp }

\maketitle\abstracts{Tempering is used to change the quark mass
while remaining in equilibrium between the trajectories of a standard hybrid
Monte Carlo simulation of four flavours of staggered fermions.  The algorithm
is faster for small enough quark masses, and particularly so when more than one
mass is required.  }
  
\section{Introduction}
The standard algorithms used for simulations of full QCD are slow, especially
for topology~\cite{us}. It is not yet possible to generate an ensemble of full
QCD configurations which samples the global topological charge adequately.

Simulated tempering~\cite{ST,us}, which traditionally promotes a parameter to a
dynamic variable that changes during the simulation, has been successfully
implemented in $\beta$ for spin glasses. We will promote the quark mass,
and change it between the trajectories of a standard hybrid
Monte Carlo $\Phi$ algorithm~\cite{hmca} for four flavours of staggered
fermions.

The choice of a new mass is made with a Metropolis step,
and insures that the change is only made if the configuration is also part of
the equilibrium distribution of the new mass. So unlike other similar
algorithms there is no need to `fix' the distribution at the end. 

The relevant measure of the speed of an algorithm is the integrated
auto-correlation~\cite{tauint}, obtained from the auto-correlation function
$C^{O}(\tau)$ for an observable $O$.  The integrated auto-correlation time
for $O$ is defined to be
\begin{equation}
\tau_{\text{int}}^{O} = \frac{1}{2} \sum_{-\infty}^{+\infty}
                                        C^{O}(\tau) 
                      \approx \frac{1}{2} \left(
                                       \frac{\sigma_{\text{max}}}{\sigma_{\text{min}}}
                                    \right)  
\label{eq:tausig}
\end{equation}
where $\sigma_{\text{max}}$ and $\sigma_{\text{min}}$ are the maximum and
minimum variances under blocking.  With too
little data, $N_{\text{data}}<1000\tau_{\text{int}}$, an accurate value for
$\tau_{\text{int}}$ cannot be obtained. An estimator is the
rightmost expression in~\eqref{eq:tausig}.

The largest value for $\tau_{\text{int}}^{O}$ over the set of all observables
$O$ defines the slowest mode, and hence the number of independent measurements.
For full QCD the global topological charge $Q$ seems
to be the slowest observable\footnote{ It may be that the topological charge
density alone is relevant for most observables in QCD. If so, slow evolution
of the global topological charge may not be such a problem. However, this
question can only be clarified once the topological sector itself has been
adequately explored.  }.

\section{Simulated Tempering}

The quark mass $\mq$ becomes a dynamic variable, and takes a new value for each 
trajectory from
a discrete, ordered set with $N_{m}$ elements, $[m_{\text{min}}, ... 
,m_{\text{max}}]$.  
This can also be seen as ranging from `slow' to `fast'. The
only requirement is that the action histograms of neighbouring masses overlap.
The simulation performs a random walk of length $N_{m}^{2}$ between the
lowest and highest masses.

The probability distribution now simulated is $P(U,\phi,i)$, the probability
of having the configuration with the set of gauge fields $U$ and
(pseudo-)fermion fields $\phi$ generated at quark mass $m_i$. For each $i$ it
is the same as the original distribution $P(U,\phi)$, of course.
This is given by
\begin{equation}
P(U,\phi,m_{i}) \propto \exp[-S(U,\phi,\beta,m_{i}) + g_{i}]
\end{equation}
where the $g_{i}$ are pre-determined constants governing the probability $P(i)$
of generating configurations at mass $m_{i}$.  This distribution is simulated
using your favourite algorithm for fixed quark mass (here HMC), combined with the
simulated tempering Metropolis
steps to change from $m_i$ to $m_{i\pm 1}$. The hybrid
Monte Carlo algorithm insures that the correct Gibbs distribution is generated
at each value of the mass, and the ST Metropolis step insures that the
configuration is also an equilibrium configuration at the new mass.

The constants $g_{i}$ can be fixed by choosing, for example, to visit each mass
with equal probability, $P(i)=1/N_{m}$. Then $g_{i} = -\ln Z_{i}$, ie. the
original free energy at fixed mass $m_{i}$. $P(i)$
is arbitrary, and can be optimized for speed. The simulation only needs
$g_{i+1}-g_{i}$, though, estimated from
\begin{equation}
\Delta g = g_{i+1}-g_{i} = - \langle\psibar\psi\rangle V\delta m
                - \langle\chi\rangle V(\delta m)^{2}
                + O((\delta m)^{3})
\label{eq:delg}
\end{equation}
where $\langle\psibar\psi\rangle$ and $\langle \chi\rangle$ are the chiral
condensate and susceptibility. The requirement of overlapping histograms
implies that $\delta m$ satisfies
\begin{equation}
\delta m \sim 1 /\sqrt{\langle\chi\rangle V} \sim m_{\sigma}/\sqrt{V}.
\end{equation}
In a short preliminary run new values for $\Delta g$ which yield a flatter
distribution can be obtained by balancing the energy difference for the changes
$i\rightarrow i+1$ and $i+1\rightarrow i$.

The overhead depends largely on $N_{m}$, which itself depends on the step size
$\delta m$.  As the susceptibility is related to the scalar meson mass, $\chi =
A_{\sigma}/m_{\sigma}^{2}$, the step size is large in the chiral limit  (at zero
temperature, but sadly not near the chiral transition).  Hence
simulated tempering becomes dramatically more effective for very small quark
masses, with the gain in speed more than compensating for the $N_{m}^{2}$
cost of having additional masses.

\section{Results}

\begin{table}[tb]
\begin{center}
\caption{Results for single mass HMC runs at $m=0.01$ and $0.02$, and simulated 
tempering between 0.01 and 0.02, and between 0.02 and 0.03. All times are in
seconds. The number in brackets indicates the time for 80 random vectors needed 
to determine $\Delta g$. The values for $\tau_{\text{int}}$ for the ST runs are 
allocated correctly! So more statistics are needed!
}
\label{tab:tauint}
\vspace{\baselineskip}
\begin{tabular*}{\textwidth}{r@{\extracolsep{\fill}}cccccc}
\hline\hline
\multicolumn{1}{c}{$m$}&
\multicolumn{1}{c}{Congrad}&
\multicolumn{1}{c}{Time}&
\multicolumn{1}{c}{ST}&
\multicolumn{1}{c}{$N_{\text{traj}}$}&
\multicolumn{1}{c}{$\tau_{\text{int}}^{Q}$}&
\multicolumn{1}{c}{Block}\\
\multicolumn{1}{c}{}&
\multicolumn{1}{c}{ratio}&
\multicolumn{1}{c}{per traj.}&
\multicolumn{1}{c}{overhead}&
\multicolumn{1}{c}{}&
\multicolumn{1}{c}{}&
\multicolumn{1}{c}{length}\\
\hline
 0.01           &      & 660 &         & 420  & 45(3) & 105 \\ 
 0.02           &      & 325 &         & 450  & 28(2) & 112 \\ 
 ST 0.01 -- 0.02 & 2.1  & 575 & 15(240) & 744  & 30(4) & 186 \\ 
 ST 0.02 -- 0.03 & 1.5  & 280 & 10(160) & 1150 & 48(4) & 105 \\ 
\hline\hline 
\end{tabular*}
\end{center}
\end{table}

We have two runs on $16^{3}\times 24$ lattices, one with six
evenly spaced masses in the range 0.01 --- 0.02, the other with nine evenly
spaced masses in the range 0.02 --- 0.03. The results obtained indicate that
seven is the optimal number. The initial estimates of $\Delta g$
from~\eqref{eq:delg} were made using data for the chiral condensate and
susceptibility from the HMC runs.  These proved adequate,
but were tuned twice to bring $P(i)$ closer to $1/N$.

We use two hybrid Monte Carlo trajectories\footnote{This is conservative, but
it insures that a configuration is seldom used for a second ST step.} at fixed
quark mass, and then a Metropolis step to change the mass. Trajectories are of
length $\tau=0.6$ with step sizes $\delta\tau=0.005$~($\mq=0.02$ and ST 0.02 ---
0.03) and $\delta\tau=0.004$~($\mq=0.01$ and ST 0.01 --- 0.02).

For four staggered fermions one only needs one extra inversion per ST step for
the transition from the set of variables $\{U,\phi,i\}$ to
$\{U,\phi,i+1\}$. This introduces negligible overhead. However, for the runs
here we have used 160 inversions (80 Gaussian noise vectors) to eliminate the
$\{\bar{\phi},\phi\}$ spread, in order to estimate the $\Delta g_{i}$ with as
few trajectories as possible.  This is not strictly correct for simulating four
staggered fermions, where $P(U,\bar{\phi},\phi)$ should be simulated. This
overhead will be larger for algorithms needing the full determinant, although
new methods for estimating the determinant~\cite{liu} may help.

\begin{figure}[bt]
  \begin{minipage}{0.48\textwidth}
    \begin{center}
      \leavevmode
     \epsfig{bbllx=65,bblly=50,bburx=230,bbury=176,file=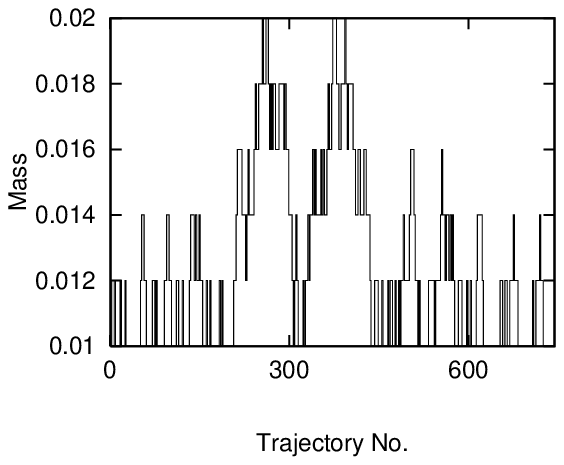}
      \caption{The time history of the mass $m_i$ for the trajectory 
	in the simulated tempering run 0.01---0.02.
        }
      \label{fig:mass}
    \end{center}
  \end{minipage}
  \hfill
  \begin{minipage}{0.48\textwidth}
    \begin{center}
      \leavevmode
        \epsfig{bbllx=65,bblly=50,bburx=230,bbury=176,file=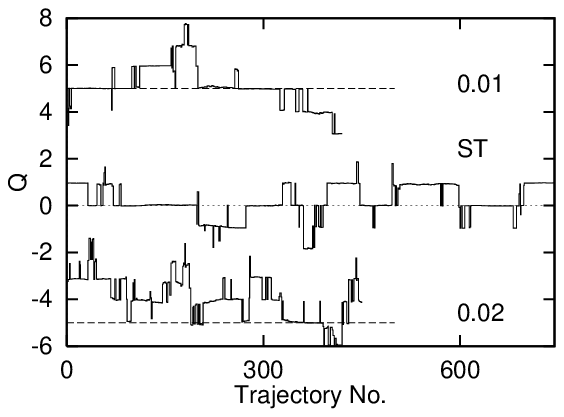}
       \caption{The time history of $Q$ from
                simulated tempering 0.01---0.02
                and pure HMC at 0.02 ($Q-5$) and 0.01 ($Q+5$).}
       \label{fig:top}
    \end{center}
  \end{minipage}
\end{figure}

The time history of the mass $m_{i}$ is shown in figure~\ref{fig:mass} for the
case 0.01---0.02. The estimates for $\Delta g$ require one further
adjustment, as there is still a bias towards the low values of the mass.
Even with relatively little tuning, though, the algorithm visits all masses in
the set.

The time history of the global topological charge, obtained as in~\cite{us}, is
shown in figure~\ref{fig:top}. There may be more movement in the topology for
the simulated tempering run than there is at fixed $m=0.01$, and less at
$m=0.02$.

In table~\ref{tab:tauint} the simulated tempering runs are compared with
standard HMC runs. The column CG ratio indicates the purely algorithmic
potential for acceleration, the ratio of the total number of conjugate gradient 
iterations needed for a complete trajectory at the smallest and largest masses
of the set.

We extract $\tau_{\text{int}}$ using~\eqref{eq:tausig} with blocking, as we do
not have anywhere near the statistics needed for an accurate estimate of
$\tau_{\text{int}}$. The true values may well be much larger, so these should
only be taken as an estimate of the lower bound. The error estimate comes from
a jack-knife analysis.

From the numbers for $\tau_{\text{int}}$ it seems plausible that the physics
moves with at least the same speed in ST, and perhaps a little faster if one is
optimistic. So less time per trajectory does mean less time for decorrelating
observables. 

The actual algorithmic acceleration can be seen in the next two columns. Since
the step size was kept constant for all masses in a set, the gain is lower than
in column CG ratio.  If only the lowest mass is needed, simulated
tempering brings little for the range 0.02---0.03, but is promising for
0.01---0.02. It would be faster, probably about 400s rather than 575s were the
step size also changed with the mass, and using the final numbers for $\Delta
g$.

Of course, the values obtained via simulated tempering for observables like the
plaquette, chiral condensate etc. agree with those from standard runs.

\section{Conclusions}

Simulated tempering does speed up full QCD simulations. It is
especially useful when going to very small masses, such as $\mq<0.01$ for
staggered fermions, as the step size goes to a constant at zero mass.
It is also more useful if results from more than one mass are required. It may
also be applied to other fermion types, eg. Wilson fermions via $\kappa$.

The conservative choice of parameters used here do not exhaust the potential
for acceleration, and leave much room for improvement.  Longer trajectories, or
one trajectory (rather than the conservative two here) between simulated
tempering steps will also reduce the overhead.  Another improvement is to run
simultaneously at each mass in the set, and then swap configurations between
adjacent masses~\cite{ST}, which implements the parallel tempering method for
fermions. This method requires considerably more memory, though, and cannot be
used if memory rather than speed limits the largest lattice used in a
simulation. Parallel tempering is under investigation, as well as an 
implementation with Wilson fermions.

\section*{Acknowledgements}
This project was partially supported by the European Union, contract
CHEX--CT92--0051, and by MURST.  GB was supported by the European Union
{\em  Human Capital and Mobility} program under HCM-Fellowship contract
ERBCHBGCT940665. The author is grateful to 
B. All\'{e}s, M. D'Elia,  A. Di Giacomo, A. Pelissetto and 
E. Vicari for valuable discussions, to B. All\'{e}s and  M. D'Elia, for 
valuable comments on the manuscript,
as well as to R. Tripiccione
for advice and assistance in using the 512 node APE/QUADRICS in Pisa.

\end{document}